\documentclass{article}
\usepackage{spconf,amsmath,amssymb,graphicx,hyperref,xcolor}
\usepackage{booktabs}
\usepackage{multirow}
\usepackage{hyperref}
\usepackage{xcolor}
\definecolor{darkgreen}{RGB}{0,128,0}
\definecolor{darkred}{RGB}{150,0,0}
\usepackage{enumitem}
\usepackage{colortbl}   
\usepackage[skip=-1pt]{caption}
\usepackage{subcaption}
\usepackage{lineno}
\usepackage{setspace}
\usepackage{float}
\usepackage{multirow}
\definecolor{bestcolor}{RGB}{255,235,156} 
\definecolor{DeepGreen}{RGB}{0, 100, 0}
\title{RECIPER: A Dual-View Retrieval Pipeline for Procedure-Oriented Materials Question Answering}
\name{Zhuoyu Wu$^{1}$, Wenhui Ou$^{2}$, Pei-Sze Tan$^{1}$, Wenqi Fang$^{3}$, Sailaja Rajanala$^{1}$, Rapha\"{e}l C.-W. Phan$^{1\dag}$
\thanks{$\dag$ Corresponding author: \texttt{raphael.phan@monash.edu} 
}
}

\address{$^1$CyPhi ($\Psi\Phi$) AI Research Lab, School of IT, Monash University, Malaysia Campus; 
\\$^2$Department of Electronic \& Computer Engineering, The Hong Kong University of Science and Technology; 
\\$^3$Shenzhen Institute of Advanced Technology, Chinese Academy of Sciences;}
\begin{document}
%
\maketitle
\begin{abstract}
Retrieving procedure-oriented evidence from materials science papers is difficult because key synthesis details are often scattered across long, context-heavy documents and are not well captured by paragraph-only dense retrieval. We present \textbf{RECIPER}, a dual-view retrieval pipeline that indexes both paragraph-level context and compact LLM-extracted procedural summaries, then combines the two candidate streams with lightweight lexical reranking. 
Across four dense retrieval backbones, RECIPER consistently improves early-rank retrieval over paragraph-only dense retrieval, achieving average gains of +3.73 in Recall@1, +2.85 in nDCG@10, and +3.13 in MRR. With \texttt{BGE-large-en-v1.5}, it reaches 86.82\%, 97.07\%, and 97.85\% on Recall@1, Recall@5, and Recall@10 respectively. We further observe improved downstream QA under automatic metrics, suggesting that procedural summaries can serve as a useful complementary retrieval signal for procedure-oriented materials QA. 

Code and data are available at \url{https://github.com/ReaganWu/RECIPER}.

\end{abstract}
\begin{keywords}
Materials science retrieval, Scientific question answering, Retrieval-augmented generation
\end{keywords}

\begin{figure*}[t]
  \centering
    \captionsetup{font={small, stretch=0.9}}
  \includegraphics[width=1.0\textwidth]{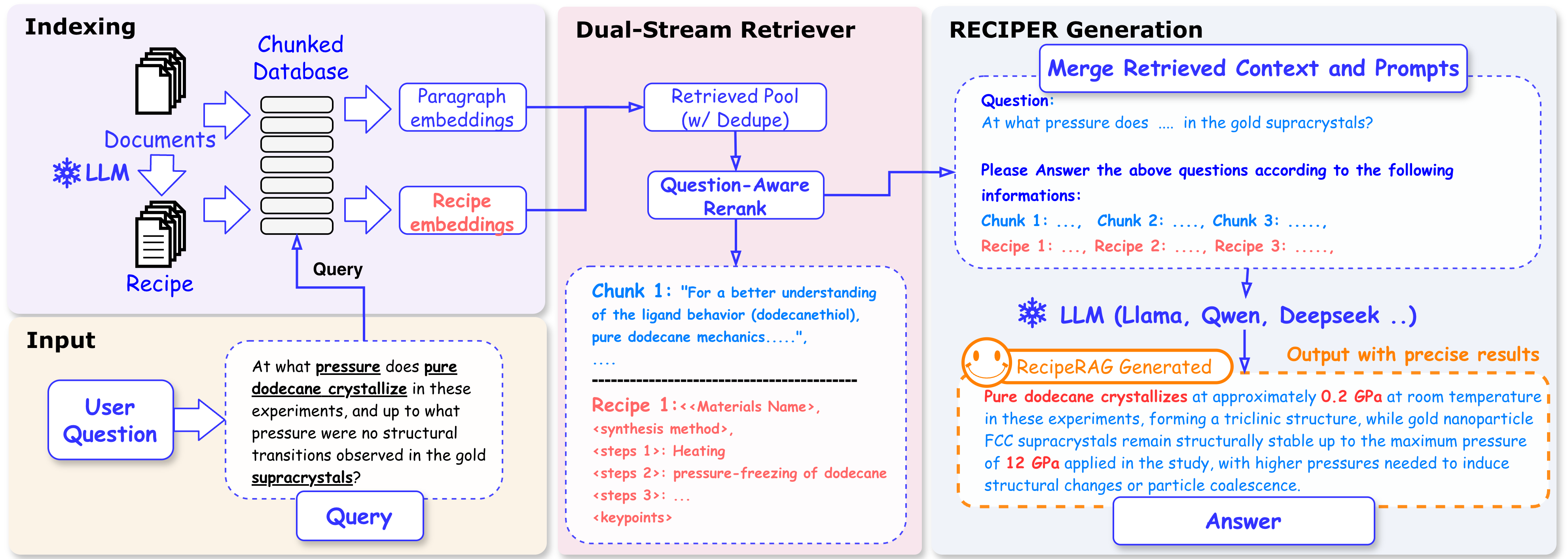}
  \caption{Overview of \textbf{RECIPER} Framework. In this framework, the user query is first transformed into feature vectors, which are then used to retrieve the most relevant \textit{Paragraph} and \textit{Recipe} embedding vectors from the \textit{Chunk Vector Database}. \textit{Recipe vectors} pass through a \textit{Threshold Screener} to filter out highly similar entries and increase content diversity. The filtered \textit{Recipes} are then combined with the \textit{Paragraphs} in a Rule-based Re-rank module and merged with the original query before being fed into the LLM, producing a precise, well-grounded answer.}
  \label{fig: general_arch}
\end{figure*}

\section{Introduction}
Large-scale scientific literature contains rich domain-specific knowledge, including experimental procedures, synthesis workflows, and contextual descriptions of materials \cite{zimmermann34examples}. However, locating such information remains labor-intensive, as key procedural details are often buried in context-heavy documents. Although large language models (LLMs) enable interactive question answering (QA) \cite{van2025surveyofAImat}, they remain unreliable for fine-grained scientific queries and may produce hallucinated responses due to their static training data \cite{zaki2023mascqa}.

Retrieval-Augmented Generation (RAG) improves reliability by retrieving supporting evidence before answer generation. Nevertheless, retrieval in materials science remains challenging. First, many questions require precise synthesis steps or material properties, whereas standard dense retrieval mainly returns unstructured text chunks, making such details difficult to identify \cite{buehler2024Melm, ansari2024agent}. Second, procedural knowledge is often distributed across multiple interdependent sections, while chunk-level retrieval breaks these connections \cite{scheidgen2023nomad}. 
Third, existing approaches have explored expert priors~\cite{ansari2024agent}, structured representations~\cite{scheidgen2023nomad}, and summary-based signals~\cite{mostafa2024g-rag}. However, it remains unclear whether compact procedural abstractions can serve as an effective auxiliary retrieval view for procedure-oriented materials QA.

To address these limitations, we propose \textbf{RECIPER}, a recipe-enhanced dual-view retrieval framework that treats procedural knowledge as a complementary retrieval signal. RECIPER represents each paper using two views: a \textit{Recipe} view, which encodes compact step-level procedural summaries, and a paragraph view, which preserves broader contextual evidence. These two views are jointly retrieved and integrated within a unified ranking pipeline, enabling more effective combination of procedural and contextual signals.

\begin{itemize}[itemsep=2pt, topsep=2pt]
    \item We introduce a dual-view retrieval pipeline that combines paragraph-level context with LLM-extracted procedural summaries for materials literature retrieval.
    \item We show empirically that procedural summaries are weak as standalone retrieval units but provide complementary signals when combined with paragraph retrieval.
    \item We demonstrate consistent gains across four dense backbones, with average improvements of +3.73 in Recall@1, +2.85 in nDCG@10, and +3.13 in MRR over paragraph-only dense retrieval, indicating that RECIPER provides a robust and backbone-agnostic improvement for scientific retrieval.
\end{itemize}


\section{Methodology}
We propose \textbf{RECIPER}, a procedure-aware dual-view retrieval framework for materials science question answering. 
The central idea is to represent each paper from two complementary views: 
(1) a \textit{contextual view} composed of paragraph-level text chunks, and 
(2) a \textit{procedural view} composed of compact LLM-extracted procedural summaries. 
Given a user query, RECIPER retrieves candidates from both views, merges them into a unified candidate pool, and applies a lightweight query-aware reranking step to prioritize evidence that is both semantically relevant and lexically aligned with the query. 
Figure~\ref{fig: general_arch} shows the overall framework.

\subsection{Procedure-Centric Knowledge Extraction}
Scientific papers in materials science often describe synthesis workflows in long, dispersed, and context-heavy paragraphs, making direct retrieval inefficient for procedure-oriented questions. 
To expose this procedural signal more explicitly, we construct a procedure-centric representation for each paper using an instruction-following LLM, DeepSeek-R1-Distill-Qwen-32B. 

For each document, the LLM generates a compact procedural summary from the full text. Each summary is formatted as a compact step-oriented description covering materials, operations, and conditions. 
Compared with raw paragraphs, these summaries compress long-form methodological text into a more retrieval-friendly form while preserving the major procedural cues needed for downstream question answering. 
We denote the set of paragraph chunks as $\mathcal{P}$ and the set of procedure-centric summaries as $\mathcal{R}$.

These summaries are not intended to replace paragraph evidence, but to expose procedural cues in a more retrieval-friendly form.

\subsection{Dual-View Candidate Retrieval}
Given a query $q$, RECIPER retrieves candidates independently from the two views. 
The paragraph view provides broad contextual evidence, while the procedural view emphasizes condensed synthesis logic and experimentally relevant operations.

Let $\mathbf{q}$ denote the query embedding, and let $\mathbf{e}_i$ denote the embedding of a candidate item from either view. 
We compute the base retrieval score as
\begin{equation}
s_i = \frac{1}{1 + d(\mathbf{q}, \mathbf{e}_i)},
\end{equation}
where $d(\cdot,\cdot)$ is the embedding-space distance. 
Using this scoring function, we retrieve the top-$K_c$ paragraph candidates
\begin{equation}
\mathcal{P}_q = \text{TopK}_{K_c}(q, \mathcal{P}),
\end{equation}
and the top-$K_c$ procedural candidates
\begin{equation}
\mathcal{R}_q = \text{TopK}_{K_c}(q, \mathcal{R}).
\end{equation}

The two candidate sets provide complementary evidence: paragraph candidates tend to preserve narrative and descriptive context, whereas procedural candidates more directly capture synthesis-oriented information.

\subsection{Candidate Merging and Stream-Aware Deduplication}
After dual-view retrieval, we merge the two candidate lists into a single pool
\begin{equation}
\mathcal{C}_q^{(0)} = \mathcal{P}_q \cup \mathcal{R}_q.
\end{equation}
Since multiple candidates may originate from the same paper and the same retrieval stream, we perform stream-aware deduplication to reduce redundant evidence while preserving cross-view complementarity. 
Specifically, for each $(\textit{paper\_id}, \textit{stream})$ pair, we keep only the highest-ranked candidate. 
This gives the filtered candidate pool
\begin{equation}
\mathcal{C}_q = \text{Dedupe}(\mathcal{C}_q^{(0)}).
\end{equation}

This design intentionally preserves cross-view complementarity while preventing within-stream redundancy from dominating the final candidate pool.

\subsection{Query-Aware Lexical Reranking}
Dense retrieval is effective for coarse semantic matching, but within a high-quality candidate pool, semantically similar candidates may still differ in how directly they address the query. 
To refine the ranking, we introduce a lightweight query-aware lexical reranking step.

Let $Q$ be the token set of the query, and let $D_i$ denote the lexical evidence of candidate $c_i$, constructed from both its title and body text, where the title provides a compact topic cue, and the body text provides local content evidence:
\begin{equation}
Q = \text{Tokenize}(q), \qquad
D_i = \text{Tokenize}(\text{title}_i \oplus \text{text}_i),
\end{equation}
where $\oplus$ denotes string concatenation. 
We define the query-coverage score of candidate $c_i$ as
\begin{equation}
o_i = \frac{|Q \cap D_i|}{\max(|Q|, 1)}.
\end{equation}
The final reranked score is then computed as
\begin{equation}
\hat{s}_i = s_i + \lambda o_i,
\end{equation}
where a small constant $\lambda$~(= 0.1) is controlling the strength of lexical adjustment.

This reranking step is intentionally lightweight. It preserves the main semantic ordering induced by dense retrieval, while promoting candidates that explicitly cover a larger fraction of the query terms. 

\subsection{Evidence Selection for Downstream QA}
Finally, all candidates in $\mathcal{C}_q$ are sorted by $\hat{s}_i$, and the top-$K$ items are selected as the evidence context
\begin{equation}
\mathcal{E}_q = \{c_1, c_2, \dots, c_K\}.
\end{equation}
The selected evidence is then passed to a downstream large language model for answer generation. 
This setup allows us to examine whether the proposed retrieval pipeline improves evidence selection and downstream answer quality under automatic metrics.

\section{Experiments}

\subsection{Experimental Setup}
We evaluate RECIPER on a materials-science QA benchmark built from 300+ research articles collected from public sources (\textit{e.g.}, arXiv and Semantic Scholar). 
Each paper is paired with GPT-5.3-generated question-answer instances and linked to its source document, yielding 1,024 query-document pairs for retrieval evaluation. 
The benchmark emphasizes synthesis-oriented questions involving procedures, material properties, and characteristic behaviors.

For retrieval, we index both paragraph chunks and procedure-centric summaries using dense embeddings. 
Unless otherwise stated, the main results use \texttt{BGE-large-en-v1.5}; we further test \texttt{all-MiniLM-L6-v2}, \texttt{Contriever}, and \texttt{E5-large-v2} to assess backbone robustness. 
We report Recall@$K$ ($K=1,5,10$), nDCG@10, and MRR. 
For downstream QA, retrieved evidence is fed into multiple LLMs ranging from 0.5B to 40B parameters, and we report BERTScore-F1, ROUGE-L, cosine similarity, and BLEURT.

\begin{table*}[htb]
\centering
\small
\setlength{\tabcolsep}{5pt}
\begin{tabular}{llccccc}
\hline
\textbf{Group} & \textbf{System} & \textbf{R@1} & \textbf{R@5} & \textbf{R@10} & \textbf{nDCG@10} & \textbf{MRR} \\
\hline

\multirow{5}{*}{External Paragraph Baselines}
& BM25~\cite{robertson2009probabilistic} & 0.6172 & 0.8066 & 0.8477 & 0.7335 & 0.6967 \\
& all-MiniLM-L6-v2~\cite{wang2020minilm} & 0.7432 & 0.9102 & 0.9307 & 0.8438 & 0.8150 \\
& Contriever~\cite{izacard2021unsupervised} & 0.7793 & 0.9131 & 0.9375 & 0.8615 & 0.8367 \\
& BGE-large-en-v1.5~\cite{bge_embedding} & 0.8408 & 0.9512 & 0.9619 & 0.9061 & 0.8875 \\
& E5-large-v2~\cite{wang2022text} & \underline{0.8477} & \underline{0.9561} & \underline{0.9717} & \underline{0.9136} & \underline{0.8945} \\
& BM25 + BGE-large-en-v1.5 & 0.7549 & 0.9443 & 0.9629 & 0.8665 & 0.8345 \\
\hline

\multirow{7}{*}{Recipe / Fusion Ablations (\textbf{BGE backbone})}
& Dense (Paragraph)~\cite{lewis2020rag} & 0.8408 & 0.9512 & 0.9619 & 0.9060 & 0.8875 \\
& Rerank (Paragraph)~\cite{sachan2022reranking} & 0.8604 & 0.9570 & 0.9619 & 0.9161 & 0.9007 \\
& Dense (Recipe)~\cite{lewis2020rag} & 0.5107 & 0.6299 & 0.6533 & 0.5837 & 0.5610 \\
& Hybrid (Recipe+Paragraph)~\cite{takahara2025materials} & 0.8486 & 0.9658 & \underline{0.9795} & 0.9181 & 0.8979 \\
& Hybrid + RRF (Recipe+Paragraph)~\cite{rackauckas2024rag} & 0.7754 & 0.9619 & 0.9707 & 0.8815 & 0.8517 \\
& Rerank (Recipe + Paragraph)~\cite{sachan2022reranking} & 0.5703 & 0.8887 & 0.9521 & 0.7634 & 0.7024 \\
& \textbf{RECIPER (Ours)} & \textbf{0.8682} & \textbf{0.9707} & \textbf{0.9785} & \textbf{0.9283} & \textbf{0.9116} \\
\hline
\end{tabular}
\caption{Retrieval performance comparison and ablation study. 
The upper block reports paragraph-based baselines, while the lower block analyzes recipe-based and dual-view variants under the BGE backbone. 
Recipe-only retrieval is weak, but combining it with paragraph retrieval improves performance, showing that procedural and contextual signals are complementary. 
RECIPER achieves the best overall results, especially on early-rank metrics, indicating more effective integration of procedural and contextual evidence.}
\label{tab:retrieval_main}
\end{table*}

\begin{table}[t]
\centering
\small
\setlength{\tabcolsep}{3.5pt}
\begin{tabular}{lcccc}
\hline
\multirow{2}{*}{\textbf{Backbone}} 
& \multicolumn{2}{c}{\textbf{vs Paragraph}} 
& \multicolumn{2}{c}{\textbf{vs Hybrid}} \\
\cline{2-5}
& $\Delta$R@1 & $\Delta$nDCG@10 & $\Delta$R@1 & $\Delta$nDCG@10 \\
\hline
MiniLM   & +0.0468 & +0.0344 & +0.0273 & +0.0151 \\
Contriever & +0.0361 & +0.0332 & +0.0263 & +0.0186 \\
BGE      & +0.0274 & +0.0223 & +0.0196 & +0.0102 \\
E5       & +0.0390 & +0.0242 & +0.0293 & +0.0138 \\
\hline
Average  & +0.0373 & +0.0285 & +0.0256 & +0.0144 \\
\hline
\end{tabular}
\caption{Cross-backbone improvement of RECIPER over paragraph-only dense retrieval and naive hybrid fusion. RECIPER consistently improves early-rank accuracy across all embedding backbones.}
\label{tab:backbone_gain}
\end{table}

\begin{table}
\centering
\small
\setlength{\tabcolsep}{3.5pt}
\begin{tabular}{lccccc}
\hline
\textbf{Model} & \textbf{M} & \textbf{BERT-F1} & \textbf{R-L} & \textbf{Cos} & \textbf{BLT} \\
\hline
GPT-5 \cite{openai2025gpt5} &$\blacklozenge$& \cellcolor{bestcolor}\textbf{0.8612} & \cellcolor{bestcolor}\textbf{0.2387} & \cellcolor{bestcolor}\textbf{0.7745} & 0.3787 \\
      &$\bullet$& 0.8552 & 0.2137 & 0.7592 & 0.3747 \\
      & $\star$ & 0.8465 & 0.2035 & 0.7350 & \cellcolor{bestcolor}{0.3794} \\
\hline
Deepseek-32B \cite{guo2025deepseek} &$\blacklozenge$& \cellcolor{bestcolor}{\textbf{0.8662}} & \cellcolor{bestcolor}{\textbf{0.2517}} & 0.7379 & \cellcolor{bestcolor}{\textbf{0.3604}} \\
             &$\bullet$& 0.8646 & 0.2428 & \cellcolor{bestcolor}{{0.7430}} & 0.3555 \\
             & $\star$ & 0.8520 & 0.1945 & 0.7030 & 0.3366 \\
\hline
Llama-3.1-8B \cite{grattafiori2024llama} &$\blacklozenge$& \cellcolor{bestcolor}{\textbf{0.8627}} & \cellcolor{bestcolor}{\textbf{0.2552}} & 0.7385 & \cellcolor{bestcolor}{\textbf{0.3615}} \\
             &$\bullet$& 0.8601 & 0.2365 & \cellcolor{bestcolor}{0.7395} & 0.3598 \\
             & $\star$ & 0.8418 & 0.1997 & 0.6668 & 0.3602 \\
\hline
Qwen-2.5-7B \cite{team2024qwen2.5} &$\blacklozenge$& \cellcolor{bestcolor}{\textbf{0.8677}} & \cellcolor{bestcolor}{\textbf{0.2752}} & \cellcolor{bestcolor}{\textbf{0.7671}} & 0.3799 \\
          &$\bullet$& 0.8655 & 0.2579 & 0.7656 & 0.3703 \\
          & $\star$ & 0.8349 & 0.1799 & 0.6377 & \cellcolor{bestcolor}{0.3925} \\
\hline
Qwen-3-4B \cite{team2024qwen2.5} &$\blacklozenge$& \cellcolor{bestcolor}{\textbf{0.8569}} &\cellcolor{bestcolor}{\textbf{0.2232}} & 0.7312 & 0.3688 \\
          &$\bullet$& 0.8553 & 0.2230 & \cellcolor{bestcolor}{0.7359} & \cellcolor{bestcolor}{0.3707} \\
          & $\star$ & 0.8339 & 0.1721 & 0.6570 & 0.3532 \\
\hline
Llama-3.2-3B \cite{grattafiori2024llama} &$\blacklozenge$& \cellcolor{bestcolor}{\textbf{0.8487}} & \cellcolor{bestcolor}{\textbf{0.2278}} & 0.7168 & \cellcolor{bestcolor}{\textbf{0.3719}} \\
             &$\bullet$& 0.8476 & 0.2121 & \cellcolor{bestcolor}{0.7197} & 0.3651 \\
             & $\star$ & 0.8409 & 0.2001 & 0.6828 & 0.3701 \\
\hline
Qwen-3-1.7B \cite{yang2025qwen3} &$\blacklozenge$& \cellcolor{bestcolor}{\textbf{0.8408}} & \cellcolor{bestcolor}{\textbf{0.2236}} & \cellcolor{bestcolor}{\textbf{0.6802}} & \cellcolor{bestcolor}{\textbf{0.4362}} \\
            &$\bullet$& 0.8341 & 0.2175 & 0.6606 & 0.4139 \\
            & $\star$ & 0.8249 & 0.1818 & 0.6283 & 0.4109 \\
\hline
Vibe-1.5B \cite{xu2025vibethinker} &$\blacklozenge$& \cellcolor{bestcolor}{\textbf{0.7626}} & \cellcolor{bestcolor}{\textbf{0.0930}} & \cellcolor{bestcolor}{\textbf{0.3361}} & \cellcolor{bestcolor}{\textbf{0.2725}} \\
          &$\bullet$& 0.7540 & 0.0809 & 0.2999 & 0.2536 \\
          & $\star$ & 0.7010 & 0.0095 & 0.0016 & 0.1986 \\
\hline
Qwen2.5-0.5B \cite{team2024qwen2.5} &$\blacklozenge$& \cellcolor{bestcolor}{\textbf{0.8524}} & \cellcolor{bestcolor}{\textbf{0.2124}} & \cellcolor{bestcolor}{\textbf{0.7532}} & \cellcolor{bestcolor}{\textbf{0.3920}} \\
             &$\bullet$& 0.8491 & 0.1966 & 0.7420 & 0.3865 \\
             & $\star$ & 0.8328 & 0.1600 & 0.6772 & 0.3844 \\
\hline
\end{tabular}
\caption{Overall QA performance across models and retrieval modes. Symbols indicate retrieval mode: $\blacklozenge$ RECIPER, $\bullet$ Paragraph-Dense RAG, $\star$ NoRAG. Metrics are BERT-F1 (F1 score of BERTScore), R-L (ROUGE-L), Cos (Cosine similarity), and BLT (BLEURT). The best value per metric is highlighted with a yellow block and in bold font.
}
\end{table}

\subsection{Retrieval Results}
Tables~\ref{tab:retrieval_main} and~\ref{tab:backbone_gain} summarize the retrieval results. 
Table~\ref{tab:retrieval_main} reports the main comparison and ablations under the BGE backbone, while Table~\ref{tab:backbone_gain} shows cross-backbone gains.

Three findings are clear from Table~\ref{tab:retrieval_main}. 
First, the procedural view alone is much weaker than paragraph-only dense retrieval, indicating that compact recipe-style summaries are insufficient as a standalone retrieval space. 
Second, naive dual-view fusion already improves over paragraph-only retrieval, confirming that contextual and procedural signals are complementary. 
Third, RECIPER further improves over naive hybrid fusion, showing that the gain comes from more effective integration of the two views.

With the BGE backbone, RECIPER improves Recall@1 from 0.8408 to 0.8682 over paragraph-only retrieval and from 0.8486 to 0.8682 over naive hybrid fusion, while also achieving the best nDCG@10 (0.9283) and MRR (0.9116). 
The improvement is most pronounced on early-rank metrics, indicating better top-evidence selection.

Table~\ref{tab:backbone_gain} shows that this trend is consistent across all four dense encoders. 
On average, RECIPER improves over paragraph-only retrieval by +3.73 points in Recall@1, +2.85 in nDCG@10, and +3.13 in MRR; compared with naive hybrid fusion, it still gains +2.56, +1.44, and +1.87 points, respectively. 
These results suggest that RECIPER is not tied to a specific embedding model, but offers a generally effective way to integrate contextual and recipe-based procedural retrieval signals.

\subsection{Transfer to Downstream QA}
We further test whether improved retrieval quality translates into better answer generation. 
Across LLMs from 0.5B to 40B parameters, RECIPER consistently outperforms both with NoRAG and Paragraph-Dense RAG on most metrics. 
The improvement is most visible on ROUGE-L and BERT-F1, suggesting that better evidence selection leads to more grounded answers. 
The effect is especially clear for smaller models, indicating that stronger retrieval can partially compensate for limited parametric knowledge.

Using Qwen-2.5-7B as an example, RECIPER improves ROUGE-L from 0.2579 to 0.2752 and BERT-F1 from 0.8655 to 0.8677 over Paragraph-Dense RAG. 
Similar trends are observed across the model spectrum, supporting that the retrieval design is architecture-agnostic and mainly benefits evidence quality rather than any specific generator.

\section{Conclusion}
In this work, we introduced \textbf{RECIPER}, a dual-view retrieval framework that integrates structured procedural knowledge with paragraph-level evidence for materials-science QA. 
Across eight LLMs ranging from 0.5B to 40B parameters, RECIPER consistently outperforms both No-RAG and paragraph-only baselines, achieving higher BERTScore, ROUGE-L, BLEURT, and semantic similarity. 
Our results show that recipe-based procedural representations complement dense retrieval by providing property- and step-level signals, with particularly strong benefits for smaller models. 
These findings indicate that RECIPER offers a robust, architecture-agnostic retrieval improvement and provides a scalable foundation for scientific QA and knowledge extraction from complex materials literature.

\vspace{-6pt}
\bibliographystyle{IEEEbib}
\bibliography{software}

\end{document}